\documentclass[11pt]{article}
\usepackage{amsfonts,amsmath,amssymb,bm,hyperref,graphicx}
\makeatletter
\newcommand\figcaption{\def\@captype{figure}\caption}
\newcommand\tabcaption{\def\@captype{table}\caption}

\makeatother \textheight 24 true cm \textwidth 16.5 true cm
\title{Associated production of a photon with dark matter pair at the
ILC within the Littlest Higgs model with T-parity}
\author{{Qing-Peng Qiao$^{1,2}$\footnote{E-mail: xxqqp@mail.nankai.edu.cn},
Bin Xu$^3$} \\
{\small \it 1. Department of Physics, Henan Normal University,
453003, Xinxiang, China} \\{\small \it 2. Department of Physics,
Henan Institute of Education, 450046, Zhengzhou, China} \\{\small
\it 3.  Department of Mathematics and Information Sciences, North
China Institute}\\{\small \it of Water Conservancy and Hydroelectric
Power, 450011, Zhengzhou, China}}
\date{}
\begin{document}
\maketitle {\bf Abstract}

Within the context of the Littlest Higgs model with T-parity, the
heavy photon ($A_{H}$) is supposed to be an ideal dark matter (DM)
candidate. One direct proof of validity of the model is to produce
the heavy photon at collider. In this paper, we investigate the
associated production of a photon with heavy photon pair at the
planned international $e^ +e^ -$ linear collider (ILC),
\textit{i.e.}, $ e^+ e^-\to A_{H}A_{H}{\gamma}$ and show the
distributions of the transverse momenta of the photon. The numerical
results indicate that the heavy photon production rate could reach
several $fb$ at the low mass parameter space and the characteristic
signal is a single high energetic photon and missing energy, carried
by the heavy photons. All in all, it can be good chance to observe
the heavy photon via this process with the high yearly luminosity of
the ILC.



\newpage

The predictions of the standard model (SM) of particle physics,
including the strong and electroweak interactions, are in accordance
with the data coming from all the experiments within the error
tolerance so far. All the facts indeed strengthen our confidence in
the SM. However, there are still some problems existing in the
framework of the SM. For example, there exits the famous hierarchy
problem. Meanwhile, SM can't provide an appropriate candidate of the
DM. These facts mean that the explanation provided by the SM should
not be the end of the story and it will let a little room for the
new physics beyond the SM. As mentioned above, the new physical
models are needed to solve the fine-tuning problem and provide the
candidates of the DM. There are many new physics models that have
been proposed, such as: Little Higgs models\cite{Cheng:2004yc,
Low:2004xc} with T-parity (LHT)\cite{Skiba:2003yf, Cheng:2003ju},
supersymmetry (SUSY) models with R-parity\cite{Dreiner:2005rd}, etc.
These models could solve the above problems successfully.

In the original little Higgs models (LHM), there is a spontaneously
broken global symmetry. The interactions of two sets cause the
explicitly broken of the global symmetry, with each set preserving
an unbroken subset. Higgs arises as an exact pseudo-Nambu-Goldstone
boson (PNGB) when either set of couplings vanishes. The hierarchy
problem is explained by introducing new heavy particles at the TeV
scale, with the identical spins of the corresponding SM particles
(opposite to the scenario in SUSY). The quadratic divergences of the
Higgs boson mass are canceled by these new particles at one-loop
level. However, the original LHM still suffer severe constraints
from precision electroweak fits that are derived from the
corrections of the tree-level interactions. To evade the issue, a
$Z_2$ discrete symmetry designated as ¡°T-parity¡±(mimic R-parity in
SUSY) is brought in. In LHT, particles of SM have the even T-parity,
while all new gauge bosons and scalar triplets have the odd
T-parity. In the above arrangement, it is completely natural to
eliminate the contributions at tree-level. Another benefit of the
LHT is that the particles with odd T-parity must be pair-produced
and then cascade decay down to the lightest T-odd particle (LTP), so
the LTP is guaranteed to be stable.

The Littlest Higgs model with T-parity \cite{Hubisz:2004ft,
Birkedal:2006fz} is such a type model. In this model, there is a
neutral LTP: heavy photon $A_H$ with an odd T-parity. If the
T-parity was conservation, it would be stable. It also has the
following characters: non-luminous, non-relativistic, non-baryonic,
and electrically neutral. \cite {Roszkowski:1991ng,
Roszkowski:1999ts, Primack:2001ia, Chen:2003bn, Ma:2004nw} The heavy
photon indeed satisfies all standards of the cold DM, so it could be
regarded as a perfect candidate of DM naturally. Experimentally
searching heavy $A_H$ gauge boson and some other special particles
\cite{Li:2011ak, Yang:2012tj} would provide direct evidence for
judging validity of the model. Because all the new particles are so
heavy that they escaped detection in prior colliders. By a general
analysis, their masses, may be in TeV regions, so that one can
expect to observe them at the high energy colliders, \textit{e.g.},
the large hadron collider (LHC) and the ILC. The production of heavy
photon pair at the LHC, $pp \to {A_H}{A_H} +X$ has been discussed in
the Ref. \cite{Hubisz:2004ft}. In fact, the LHC might be difficult
to confirm a theory, but feasible to rule out a model as long as the
signal predicted by the model does not show up at the required
region. If we want to better understand the properties of the
Littlest Higgs model with T-parity, we need to study the production
of these particles at the ILC because of the clean background. In
our previous work, we have studied $ e^+ e^ - \to A_{H}A_{H}$\cite
{Qiao:2011kp} at the ILC in the Littlest Higgs model with T-parity.
However, because the interactions between SM particles and heavy
photons are weakness, the detection of above process is difficult.
In this work, we analyze associated production of a photon with dark
matter pair at the ILC, that is, $ e^+ e^ - \to A_{H}A_{H}
{\gamma}$, where typical collider signals are a single high
energetic photon and missing energy, carried by the heavy photons.

This paper is arranged as follows. The numerical results of the
production rate and the distributions of the transverse momenta are
given in Sec. II where all input parameters are explicitly listed.
Our analysis of various aspects and conclusions are presented in the
final section.

\section{Theoretical formulation and numerical results}

~~~~We know the Littlest Higgs model with T-parity is based on a
non-linear $\sigma$-model describing a global $SU(5)/SO(5)$ symmetry
breaking, which takes place at an energy scale $\Lambda\sim4\pi
f\sim$10 TeV. The vacuum expectation value (VEV) that causes the
breaking is characterized by the direction of the form

\begin{eqnarray}
{\Sigma _0} = \left( {\begin{array}{*{20}{c}}
   {} & {} & {} & 1 & {}  \\
   {} & {} & {} & {} & 1  \\
   {} & {} & 1 & {} & {}  \\
   1 & {} & {} & {} & {}  \\
   {} & 1 & {} & {} & {}  \\
\end{array}} \right).
\end{eqnarray}

The vacuum also breaks the assumed embedded local gauge symmetry
$[SU(2)\times U(1)]^2$ subgroup down to the diagonal subgroup. One
set of $SU(2)\times U(1)$ acquire masses of order $f$
\begin{eqnarray}
 W_H^\alpha  = \frac{1}{{\sqrt 2 }}(W_1^\alpha  - W_2^\alpha ),~~~~{M_{W_H^\alpha }} = gf, \\
 {B_H} = \frac{1}{{\sqrt 2 }}({B_1} - {B_2}),~~~~{M_{{B_H}}} = \frac{{g'}}{{\sqrt 5 }}f,
\end{eqnarray}
the other set remaining massless before Electroweak Symmetry
Breaking (EWSB) is identified as the SM electroweak gauge fields
$SU(2)_L\times U(1)_Y$.

After EWSB, the VEV of the Higgs will shift the mass eigenstates of
the heavy gauge boson sector. The new heavy mass eigenstates $A_H$,
$Z_H$ and ${W_H}^\pm$ could be written as
\begin{eqnarray}
\begin{array}{l}
 W_H^ \pm  = \frac{\displaystyle 1}{\displaystyle{\sqrt 2 }}(W_H^1 \mp iW_H^2), \\
 {Z_H} = \sin {\theta _H}{B_H} + \cos {\theta _H}W_H^3, \\
 {A_H} = \cos {\theta _H}{B_H} - \sin {\theta _H}W_H^3, \\
 \end{array}
\end{eqnarray}
the masses of the new heavy gauge bosons receive corrections of
order $v^2/f^2$ and could be written as
\begin{flalign}
 \begin{split}
M(Z_H)=M(W_H^\pm)=gf(1 - \frac{{{v^2}}}{{8{f^2}}}) \approx 0.65f,\\
~{\rm{ }}M({A_H})= \frac{{fg'}}{{\sqrt 5 }}(1 -
\frac{{5{v^2}}}{{8{f^2}}}) \approx 0.16f,
 \end{split}
 \end{flalign}
where $g$ and $g'$ are the gauge couplings of the SM $SU(2)_L$ and
$U(1)_Y$ respectively. So the mass of the $A_H$ is distinctively
smaller than other T-odd particles, which are generically of level
$f$. With the T-parity, $A_H$ is a weakly interacting stable neutral
particle. It provides a natural candidate of the weakly-interacting
massive particle (WIMP)\cite{Bertone:2004pz} cold DM.

The mirror fermions, acquire masses through a Yukawa-type
interaction
\begin{eqnarray}
\kappa f({\bar \Psi _2}\xi \Psi ' + {\bar \Psi _1}{\Sigma _0}\Omega
{\xi ^\dag }\Omega \Psi ')
\end{eqnarray}
whereas $\Psi _1$, $\Psi _2$ are the fermion doublets and $\Psi '$
is a doublet under $SU(2)_2$.

One fermion doublet ${\Psi _H} = \frac{1}{{\sqrt 2 }}({\Psi _1} +
{\Psi _2})$ gets a mass $\kappa f$, $f$ is a scale parameter and
takes the value of 1000. Concretely, the T-odd heavy partners of the
SM leptons get the following masses $\sqrt{2} \kappa_l f$
\cite{Cacciapaglia:2009cv}, where the $\kappa_l$ is the independent
Yukawa coupling of flavor. The T-odd  heavy lepton will be supposed
to surpass 300 GeV to evade the colored T-odd particles from being
detected in the squark searches at the Tevatron.

In this model, the coupling terms that heavy photon interact with SM
fermion and T-odd fermion, the SM photon interact with T-odd
fermions are shown as:\cite{Hubisz:2004ft, Birkedal:2006fz}
\begin{eqnarray}
 \begin{array}{l}
A_H^\mu {\tilde L_i}{L_j}: ~i\frac{e}{{10{C_W}{S_W}}}({S_W} -
5{C_W}{(\frac{v}{f})^2}{x_h}){\gamma ^\mu }{P_L}\delta_{ij},\\
A^\mu {\tilde L_i}{\tilde L_j}:~ie{\gamma ^\mu }{\delta _{ij}},
 \end{array}
\end{eqnarray}
where $\tilde L$ is the heavy T-odd lepton and $L$ is the SM lepton.
$e=\sqrt{4\pi\alpha}$, where $\alpha$ is the fine-structure constant
and take the value\cite{Nakamura:2010zzi} of 1/128. ${x_h} =
\frac{5}{4}\frac{{gg'}}{{5{g^2} - {{g'}^2}}}$, $v =
\frac{{2{M_W}{S_W}}}{e}$,  ${\rm{ }}{P_L} = \frac{ \displaystyle {1
- {\gamma_5}}}{\displaystyle 2}$ is the left-handed chiral
projection operator. $S_W$ and $C_W$ are sine and cosine of the
Weinberg angle respectively, $M_W$ is the mass of SM W gauge boson.

\begin{figure}
\centering
\includegraphics[height=4.5cm,angle=0]{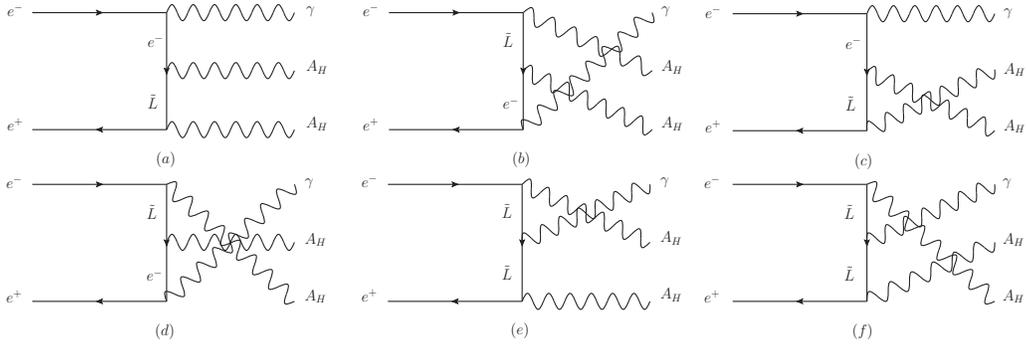} \caption{The Feynman
diagrams of the process $e^ + e^ - \to A_{H}A_{H}{\gamma}$.}
\label{fig.1} \end{figure}

At the tree-level, the Feynman graphs for the process of $e^ + e^ -
\to A_{H}A_{H}{\gamma}$  are shown in Fig. \ref{fig.1}.

According to the Feynman graphs in Fig. \ref{fig.1}, we can directly
write the explicit amplitude of the process:
\begin{eqnarray}\label{eqnarray}
\begin{array}{l}
M = {M_a} + {M_b}+{M_c}+{M_d}+{M_e}+{M_f},\\
{M_a} =  - i\frac{{{e^3}}}{{{{(10{S_W}{C_W})}^2}}}{({S_W} -
5{C_W}{(\frac{v}{f})^2}{x_h})^2}\frac{1}{{P_{5M2}^2 - m_{\tilde
L}^2}}\frac{1}{{P_{1M3}^2 - m_e^2}}{\overline v _{{e^ +
}}}({P_2}){\gamma ^\rho }({{\rlap{$\not$} P}_{5M2}} + {m_{\tilde
L}}){\gamma ^\nu } {\rm{         }}{P_L}\\~~~~~~~~({{\rlap{$\not$}
P}_{1M3}} + {m_e}){\gamma ^\mu }{P_L}
{u_{{e^ - }}}({P_1}){\epsilon _\mu }({P_3}){\epsilon _\nu }({P_4}){\epsilon _\rho }({P_5}), \\
 {M_b} =  - i\frac{{{e^3}}}{{{{(10{S_W}{C_W})}^2}}}{({S_W} - 5{C_W}{(\frac{v}{f})^2}{x_h})^2}\frac{1}{{P_{3M2}^2 - m_{\tilde L}^2}}\frac{1}{{P_{1M4}^2 - m_e^2}}{\overline v _{{e^ + }}}({P_2}){\gamma ^\mu }({{\rlap{$\not$} P}_{3M2}} + {m_{\tilde L}}){\gamma ^\rho }{\rm{         }}{P_L}\\~~~~~~~~({{\rlap{$\not$} P}_{1M4}} + {m_e}){\gamma ^\nu }{P_L}
{u_{{e^ - }}}({P_1}){\epsilon _\mu }({P_3}){\epsilon _\nu }({P_4}){\epsilon _\rho }({P_5}), \\
 {M_c} =  - i\frac{{{e^3}}}{{{{(10{S_W}{C_W})}^2}}}{({S_W} - 5{C_W}{(\frac{v}{f})^2}{x_h})^2}\frac{1}{{P_{4M2}^2 - m_{\tilde L}^2}}\frac{1}{{P_{1M3}^2 - m_e^2}}{\overline v _{{e^ + }}}({P_2}){\gamma ^\nu }({{\rlap{$\not$} P}_{4M2}} + {m_{\tilde L}}){\gamma ^\rho }{\rm{         }}{P_L}\\~~~~~~~~({{\rlap{$\not$} P}_{1M3}} + {m_e}){\gamma ^\mu }{P_L}
{u_{{e^ - }}}({P_1}){\epsilon _\mu }({P_3}){\epsilon _\nu }({P_4}){\epsilon _\rho }({P_5}), \\
 {M_d} =  - i\frac{{{e^3}}}{{{{(10{S_W}{C_W})}^2}}}{({S_W} - 5{C_W}{(\frac{v}{f})^2}{x_h})^2}\frac{1}{{P_{3M2}^2 - m_{\tilde L}^2}}\frac{1}{{P_{1M5}^2 - m_e^2}}{\overline v _{{e^ + }}}({P_2}){\gamma ^\mu }({{\rlap{$\not$} P}_{3M2}} + {m_{\tilde L}}){\gamma ^\nu }{\rm{         }}{P_L}\\~~~~~~~~({{\rlap{$\not$} P}_{1M5}} + {m_e}){\gamma ^\rho }{P_L}
{u_{{e^ - }}}({P_1}){\epsilon _\mu }({P_3}){\epsilon_\nu }({P_4}){\epsilon _\rho }({P_5}), \\
{M_e} =  - i\frac{{{e^3}}}{{{{(10{S_W}{C_W})}^2}}}{({S_W} -
5{C_W}{(\frac{v}{f})^2}{x_h})^2}\frac{1}{{P_{5M2}^2 - m_{\tilde
L}^2}}\frac{1}{{P_{1M4}^2 - m_{\tilde L}^2}}{\overline v _{{e^ +
}}}({P_2}){\gamma ^\rho }{P_L}({{\rlap{$\not$} P}_{5M2}} +
{m_{\tilde L}}){\gamma ^\mu }\\~~~~~~~~ {\rm{
}}({{\rlap{$\not$} P}_{1M4}} + {m_{\tilde L}}){\gamma ^\nu }{P_L}
{u_{{e^ - }}}({P_1}){\epsilon _\mu }({P_3}){\epsilon_\nu }({P_4}){\epsilon _\rho }({P_5}), \\
{M_f} =  - i\frac{{{e^3}}}{{{{(10{S_W}{C_W})}^2}}}{({S_W} -
5{C_W}{(\frac{v}{f})^2}{x_h})^2}\frac{1}{{P_{4M2}^2 - m_{\tilde
L}^2}}\frac{1}{{P_{1M5}^2 - m_{\tilde L}^2}}{\overline v _{{e^ +
}}}({P_2}){\gamma ^\nu}{P_L}({{\rlap{$\not$} P}_{4M2}} + {m_{\tilde
L}}){\gamma ^\mu }\\~~~~~~~~ {\rm{         }}({{\rlap{$\not$}
P}_{1M5}} + {m_{\tilde L}}){\gamma ^\rho }{P_L}{u_{{e^ -
}}}({P_1}){\epsilon _\mu }({P_3}){\epsilon_\nu }({P_4}){\epsilon
_\rho }({P_5}),
\end{array}
\end{eqnarray}
where $P_{5M2}=P_5-P_2$, $P_{1M3}=P_1-P_3$, etc. The $\epsilon$s are
the polarization vectors for the final gauge bosons.

With the above production amplitude, we can obtain the production
cross section of the process and the distributions of the transverse
momenta of $A_{H}$ through the Monte Carlo method.

\begin{figure}
\centering
\includegraphics[height=7cm,angle=0]{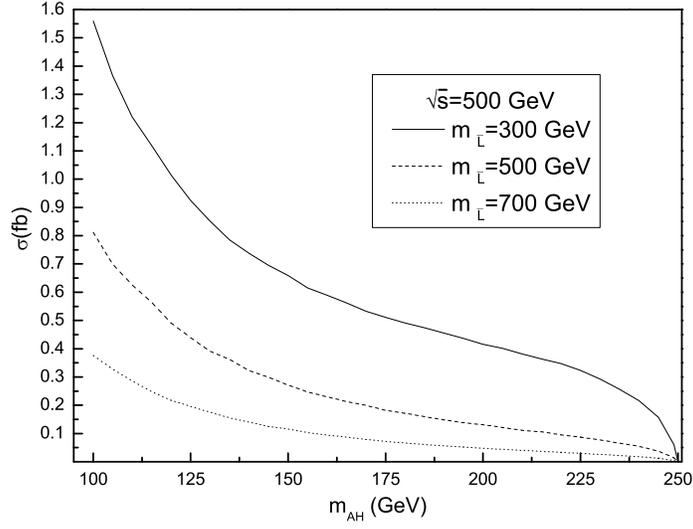} \caption{The dependence of the cross section of $ e^ + e^ - \to
A_{H}A_{H}{\gamma}$ on heavy photon mass $m_{AH}$ (100$\sim $250
GeV) for $\sqrt{s}$=500 GeV and  $m_{\tilde{L}}=$300, 500, 700 GeV
at the ILC.} \label{fig.2} \end{figure}

\begin{figure}
\centering
\includegraphics[height=7cm,angle=0]{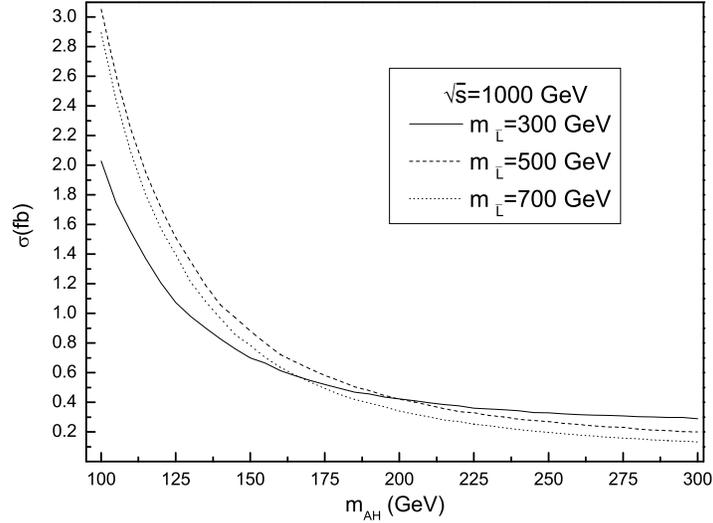} \caption{The dependence of the cross section of $ e^ + e^ - \to
A_{H}A_{H}{\gamma}$ on heavy photon mass $m_{AH} $ (100$\sim $300
GeV) for $\sqrt{s}$=1000 GeV and  $m_{\tilde{L}}$=300, 500, 700 GeV
at the ILC.} \label{fig.3} \end{figure}

There are also three free parameters involved in the production
amplitudes: the energy of the center-of-mass frame $\sqrt{s}$, the
heavy photon mass $m_{AH}$ and the mass of heavy lepton
$m_{\tilde{L}}$. Concretely, in order to expose possible dependence
of the cross section on these parameters, we take two groups of
values: $\sqrt{s}=$500, 1000 GeV, $m_{\tilde{L}}= $300, 500, 700 GeV
respectively, we also let $m_{AH}$ vary in the ranges 100 to 250 GeV
in Fig. \ref{fig.2} and 100 to 300 GeV in Fig. \ref{fig.3}. The
final numerical results of the cross sections are illustrated in
Fig. \ref{fig.2} and Fig. \ref{fig.3}.

From these figures, we can observe that the dependence of production
rate on heavy photon mass is quite strong, the cross section
magnitude has a slide of above one order when $m_{AH}$ arises from
100 GeV to 225 GeV,  naturally since the phase space is depressed
rigorously by large $m_{AH}$. The dependence of the cross section on
$\sqrt{s}$ is obvious: when $\sqrt{s}$ becomes large, the cross
section increases evidently. The relations between the cross section
and $m_{\tilde{L}}$, on the other hand, have a little complications,
when $\sqrt{s}$ equals 500 GeV, $m_{\tilde{L}}$ becomes large, the
cross section decreases, but when $\sqrt{s}$ equals 1000 GeV, that
is no longer the case, specific relations can be obtained from Fig.
\ref{fig.3}. We also find that the production rate of $ e^ + e^ -
\to A_{H}A_{H}{\gamma}$ is much large than $ e^ + e^ - \to
A_{H}A_{H}$\cite{Qiao:2011kp} of the same parameters. Despite
depressions from the three-body phase space and the $\tilde Y$
factor, the extra items which contain $m_{\tilde L}$ in the
numerators of the Eq. (\ref {eqnarray}) enhance the final results
strongly.

\begin{figure}
\centering
\includegraphics[height=6.5cm]{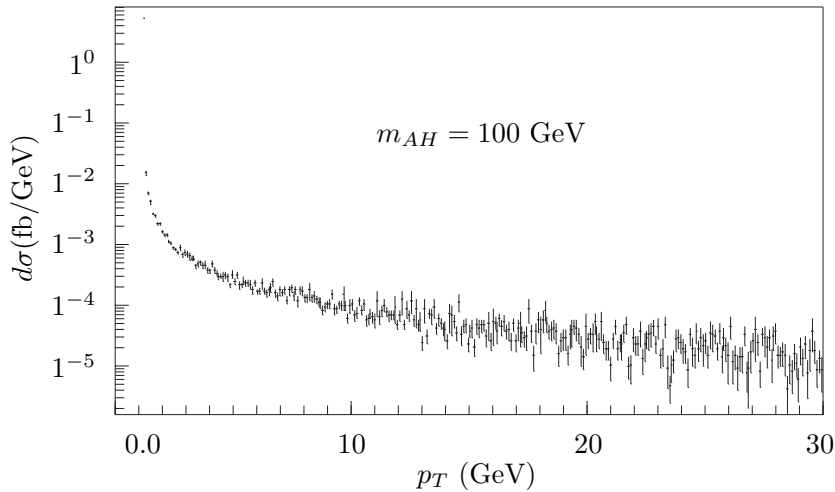} \caption{The
distributions of the transverse momenta of final photon for the
process $ e^ + e^ - \to A_{H}A_{H}{\gamma}$ for $\sqrt{s}$=500 GeV
and $m_{AH}$=100 GeV.} \label{fig.4}
\end{figure}

\begin{figure}
\centering
\includegraphics[height=6.5cm]{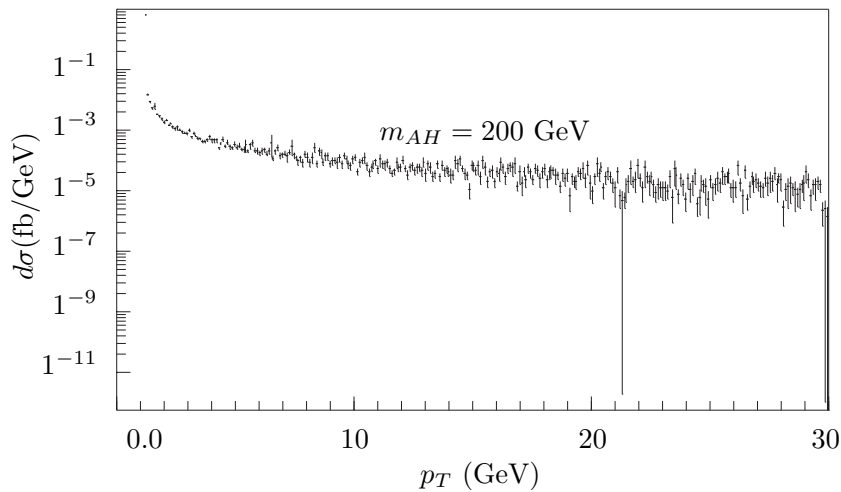} \caption{The
distributions of the transverse momenta of final photon for the
process $ e^ + e^ - \to A_{H}A_{H}{\gamma}$ for $\sqrt{s}$=500 GeV
and $m_{AH}$=200 GeV.} \label{fig.5}
\end{figure}

In Fig. \ref{fig.4} and Fig. \ref{fig.5}, with $\sqrt{s}$=500 GeV,
we present the transverse momentum distributions of final photon for
$m_{AH}$=100 and 200 GeV, respectively. We can find that the
differential cross section with the small transverse momentum of
final photon make the main contribution to the cross section of $ e^
+ e^ - \to A_{H}A_{H}{\gamma}$.

\section{Discussions and conclusions}

~~~~The combination of cosmology and high energy collider
\cite{Khlopov:1998ux, Belotsky:1998pv} seems to be a good idea.
Astrophysical observations provide a way to study the
characteristics of DM, however, the exact properties of a DM
particle need to be determined by the DM factories, that is to say,
the LHC and the ILC. Because of relatively clean background, the ILC
is more appropriate for accurate examinations. So this paper is a
proper supplement to Refs. \cite{Hubisz:2004ft, Qiao:2007rc}.
Finally, the heavy photon is the LTP and must be produced in pairs,
constraint from the final state phase space of $A_{H}A_{H}{\gamma}$
is alleviated compared with other heavy T-odd particles pair
production, since heavier T-odd particles are too heavy to be
pair-produced in the first stage of the ILC with $\sqrt{s}$ =500
GeV. It may imply that if the Littlest Higgs model with T-parity
applies, $A_{H}A_{H}{\gamma}$ would be detected at earliest time at
the ILC.

One factor strongly complicates the investigation of dark matter in
the collider experiments is that DM particles do not carry either
electric or color charge. In the collider experiments, the direct
detection of DM at a cllider is very difficult, DM would be like
neutrinos and therein they would escape the detector without
depositing energy in the experimental devices, causing an obvious
imbalance of momentum and energy in collider events. In this paper,
unlikely the radiative production of heavy photons in  $e^+e^-$
annihilation of six-dimensional SUSY QED\cite{Fayet:1985kt}, the
signal of the heavy photon pair can be observed as missing momentum
recoiling against the detected photon by measuring the energy
deposited in each calorimeter cell of a detector. In the SM, the
main irreducible background originates from the reaction ${e^ + }{e^
- } \to \nu \bar \nu \gamma$. This reaction is dominated by the
t-channel W exchange contribution at the energies well above the Z
peak. The background cross section at the ILC ($\sqrt{S}$=500 GeV)
with no polarization was provided in Ref. \cite{Birkedal:2004xn},
which is treated as the function of the DM mass, and has a rather
large value. Nevertheless, the rate predicted of $ e^ + e^ - \to
A_{H}A_{H}{\gamma}$ may well be observable. Experiments at the ILC
can search for $A_{H}A_{H}{\gamma}$ signature as an excess over the
SM background $ \nu \bar \nu \gamma$. If all of the background
caused by SM neutrinos and the uncertainties were been subtracted
and the vector sum of all the transverse momenta still not equal to
zero, we could believe that something invisible is produced, the
undetected particle(s) may be the DM candidate(s) (such as a heavy
photon).

The second problem we have to solve is how to distinguish the heavy
photon of the Littlest Higgs model with T-parity from other type DM
production at the ILC, \textit{e.g.}, the lightest neutralino
($\tilde \chi _1^0$) of SUSY with R-parity, the gravitino of
SUSY\cite{Bolz:2000fu}, the heavy neutrinos of new
generations\cite{Fargion:1999ss}, etc. Since different DM particles
have different reaction channels and probabilities to be detected by
the detectors of the ILC, which offer a way to distinguish the $A_H$
from other DM candidates. Tree-level production of ${e^ + }{e^ -
}\to \tilde \chi _1^0  \tilde \chi_1^0
{\gamma}$\cite{Dreiner:2006sb} in the MSSM\cite{Goldberg:1983nd,
Ellis:1983ew} with R-parity, ${e^ + }{e^ - } \to {\Psi _g}{\bar \Psi
_g}\gamma$ within SUSY\cite{Fayet:1982ky}, ${e^ + }{e^ - } \to \nu
\bar \nu \gamma$\cite{Fargion:1995qb} of the new generations have
been studied in detail in the relative papers respectively. The
other way to the model discrimination is by detecting and measuring
the shape of the photon spectrum in the events with WIMP
production.\cite{Konar:2009ae} Readers who are interested in and
want to know more about the issue can see these papers and we don't
discuss the problem more.

As a conclusion, our calculations indicate that the production rate
of $e^+ e^- \to A_{H}A_{H}{\gamma}$ could reach several $fb$ in the
relatively low mass parts of the allowed parameter space. That is to
say, thousands of signal events would be produced one year via the
$A_{H}A_{H}{\gamma}$ production mode at the ILC thinks to its high
energy and yearly luminosity (500 $fb^{-1}$ at 500 GeV first and
1000 $fb^{-1}$ at 1000 GeV later). The advantage of analyzing such
processes at the ILC is obvious because the hadronic background is
very suppressed and the amount of signals may be practically
observable. Therefore, associated production of a photon with heavy
photon pair at the ILC might be a promising signal for the Littlest
Higgs model with T-parity and a proper way to detect the DM.

\section*{Acknowledgments}

~~~~This work is supported by the National Natural Science
Foundation of China (No. 11075045) and the Natural Science
Foundation of Education Department of Henan Province (No.
2011A140005).


\newpage
\begin{thebibliography}{a}\vspace{5mm}

\bibitem{Cheng:2004yc}
  H.~C.~Cheng and I.~Low,
  JHEP {\bf 0408}, 061 (2004)
  [arXiv:hep-ph/0405243].

\bibitem{Low:2004xc}
  I.~Low,
  JHEP {\bf 0410}, 067 (2004)
  [arXiv:hep-ph/0409025].

\bibitem{Skiba:2003yf}
  W.~Skiba and J.~Terning,
  Phys.\ Rev.\  D {\bf 68}, 075001 (2003)
  [arXiv:hep-ph/0305302].

\bibitem{Cheng:2003ju}
  H.~C.~Cheng and I.~Low,
  JHEP {\bf 0309}, 051 (2003)
  [arXiv:hep-ph/0308199].

\bibitem{Dreiner:2005rd}
  H.~K.~Dreiner, C.~Luhn and M.~Thormeier,
  Phys.\ Rev.\  D {\bf 73}, 075007 (2006)
  [arXiv:hep-ph/0512163].

\bibitem{Hubisz:2004ft}
  J.~Hubisz and P.~Meade,
  Phys.\ Rev.\  D {\bf 71}, 035016 (2005)
  [arXiv:hep-ph/0411264].

\bibitem{Birkedal:2006fz}
  A.~Birkedal, A.~Noble, M.~Perelstein and A.~Spray,
  Phys.\ Rev.\  D {\bf 74}, 035002 (2006)
  [arXiv:hep-ph/0603077].

\bibitem{Roszkowski:1991ng}
  L.~Roszkowski,
  Phys.\ Lett.\  B {\bf 262}, 59 (1991).

\bibitem{Roszkowski:1999ts}
  L.~Roszkowski,
  arXiv:hep-ph/9903467.

\bibitem{Primack:2001ia}
  J.~R.~Primack,
  arXiv:astro-ph/0112255.

\bibitem{Chen:2003bn}
  P.~Chen,
  arXiv:astro-ph/0303349.

\bibitem{Ma:2004nw}
  C.~P.~Ma and M.~Boylan-Kolchin,
  Phys.\ Rev.\ Lett.\  {\bf 93}, 021301 (2004)
  [arXiv:astro-ph/0403102].

\bibitem{Li:2011ak}
  B.~-Z.~Li, J.~-Z.~Han and B.~-F.~Yang,
  Commun.\ Theor.\ Phys.\  {\bf 56}, 703 (2011).

\bibitem{Yang:2012tj}
  B.~Yang,
  Commun.\ Theor.\ Phys.\  {\bf 57}, 849 (2012)
  [arXiv:1204.0845 [hep-ph]].

\bibitem{Qiao:2011kp}
  Q.~P.~Qiao and B.~Xu,
  arXiv:1105.3555 [hep-ph].

\bibitem{Bertone:2004pz}
  G.~Bertone, D.~Hooper and J.~Silk,
  Phys.\ Rept.\  {\bf 405}, 279 (2005)
  [arXiv:hep-ph/0404175].

\bibitem{Cacciapaglia:2009cv}
  G.~Cacciapaglia, A.~Deandrea, S.~R.~Choudhury and N.~Gaur,
  Phys.\ Rev.\ D {\bf 81}, 075005 (2010)
  [arXiv:0911.4632 [hep-ph]].

\bibitem{Nakamura:2010zzi}
  K.~Nakamura {\it et al.}  [Particle Data Group],
  J.\ Phys.\ G {\bf 37}, 075021 (2010).

\bibitem{Khlopov:1998ux}
  M.~Y.~Khlopov, A.~S.~Sakharov and A.~L.~Sudarikov,
  Grav.\ Cosmol.\  {\bf 4}, S1 (1998)
  [arXiv:hep-ph/9811432].

\bibitem{Belotsky:1998pv}
  K.~M.~Belotsky, M.~Y.~Khlopov, A.~S.~Sakharov, A.~A.~Shklyaev and A.~L.~Sudarikov,
  Grav.\ Cosmol.\ Suppl.\  {\bf 4} (1998) 70.

\bibitem{Qiao:2007rc}
  Q.~P.~Qiao, J.~Tang and X.~Q.~Li,
  Commun.\ Theor.\ Phys.\  {\bf 50}, 1211 (2008)
  [arXiv:0711.3254 [hep-ph]].

\bibitem{Fayet:1985kt}
  P.~Fayet,
  Nucl.\ Phys.\  B {\bf 263}, 649 (1986).

\bibitem{Birkedal:2004xn}
  A.~Birkedal, K.~Matchev and M.~Perelstein,
  Phys.\ Rev.\  D {\bf 70}, 077701 (2004)
  [arXiv:hep-ph/0403004].

\bibitem{Bolz:2000fu}
  M.~Bolz, A.~Brandenburg and W.~Buchmuller,
  Nucl.\ Phys.\  B {\bf 606}, 518 (2001)
  [Erratum-ibid.\  B {\bf 790}, 336 (2008)]
  [arXiv:hep-ph/0012052].

\bibitem{Fargion:1999ss}
  D.~Fargion, Yu.~A.~Golubkov, M.~Y.~Khlopov, R.~V.~Konoplich and R.~Mignani,
  Pisma Zh.\ Eksp.\ Teor.\ Fiz.\  {\bf 69}, 402 (1999)
  [JETP Lett.\  {\bf 69}, 434 (1999)]
  [arXiv:astro-ph/9903086].

\bibitem{Dreiner:2006sb}
  H.~K.~Dreiner, O.~Kittel and U.~Langenfeld,
  Phys.\ Rev.\  D {\bf 74}, 115010 (2006)
  [arXiv:hep-ph/0610020].

\bibitem{Goldberg:1983nd}
  H.~Goldberg,
  Phys.\ Rev.\ Lett.\  {\bf 50}, 1419 (1983)
  [Erratum-ibid.\  {\bf 103}, 099905 (2009)].

\bibitem{Ellis:1983ew}
  J.~R.~Ellis, J.~S.~Hagelin, D.~V.~Nanopoulos, K.~A.~Olive and M.~Srednicki,
  Nucl.\ Phys.\  B {\bf 238}, 453 (1984).

\bibitem{Fayet:1982ky}
  P.~Fayet,
  Phys.\ Lett.\  B {\bf 117}, 460 (1982).

\bibitem{Fargion:1995qb}
  D.~Fargion, M.~Y.~Khlopov, R.~V.~Konoplich and R.~Mignani,
  Phys.\ Rev.\  D {\bf 54}, 4684 (1996).

\bibitem{Konar:2009ae}
  P.~Konar, K.~Kong, K.~T.~Matchev and M.~Perelstein,
  New J.\ Phys.\  {\bf 11}, 105004 (2009)
  [arXiv:0902.2000 [hep-ph]].

\end{thebibliography}
\end{document}